\documentclass[pra,twocolumn,nofootinbib,showpacs,superscriptaddress]{revtex4-1}

\usepackage{graphicx}
\usepackage[colorlinks=true,citecolor=blue,urlcolor=black]{hyperref}
\usepackage{amsfonts}
\usepackage{amsmath,amssymb}
\usepackage{dsfont}
\usepackage{euscript}
\usepackage{float}
\usepackage{amsthm}
\usepackage{mathalfa}
\usepackage{bbm}
\usepackage{ragged2e}
\usepackage{color}
\usepackage{xcolor}
\definecolor{shadecolor}{RGB}{150,150,150}

\usepackage{epstopdf}
\usepackage{epsfig}
\usepackage{mdframed}
\newcommand{\tn}[1]{\textnormal{#1}}
\newcommand{\rv}[1]{\hat{#1}}

\usepackage{cleveref}

\def\be{\begin{equation}}
\def\ee{\end{equation}}
\newcommand{\Tr}{\mathrm{ Tr }}
\newcommand{\reals}{\ensuremath{\mathbb{R}}}

\newcommand{\sket}[1]{{\ensuremath{\lvert#1\rangle}}}
\newcommand{\lket}[1]{{\ensuremath{\left\lvert#1\right\rangle}}}
\newcommand{\ket}[1]{\if@display\lket{#1}\else\sket{#1}\fi}

\newcommand{\sbra}[1]{{\ensuremath{\langle#1\rvert}}}
\newcommand{\lbra}[1]{{\ensuremath{\left\langle#1\right\rvert}}}
\newcommand{\bra}[1]{\if@display\lbra{#1}\else\sbra{#1}\fi}

\newcommand{\sbraket}[2]{{\ensuremath{\langle#1\rvert#2\rangle}}}
\newcommand{\lbraket}[2]{{\ensuremath{\left\langle#1\!\left\rvert\vphantom{#1}#2\right.\!\right\rangle}}}
\newcommand{\braket}[2]{\if@display\lbraket{#1}{#2}\else\sbraket{#1}{#2}\fi}

\newcommand{\sketbra}[2]{{\ensuremath{\lvert #1\rangle\!\langle #2\rvert}}}
\newcommand{\lketbra}[2]{{\ensuremath{\left\lvert #1\right\rangle\!\!\left\langle #2\right\rvert}}}
\newcommand{\ketbra}[2]{\if@display\lketbra{#1}{#2}\else\sketbra{#1}{#2}\fi}

\newcommand{\rs}{{\rm{1}}}
\newcommand{\rd}{{\rm{2}}}
\newcommand{\rdd}{{\rm{3}}}

\newcommand{\eps}{\epsilon}

\theoremstyle{dotless}
\newtheorem{thm}{Theorem}
\newtheorem{theorem}{Theorem}

\theoremstyle{definition}
\newtheorem{definition}[theorem]{Definition}

\newcommand{\thistheoremname}{}
\newtheorem{genericthm}[thm]{\thistheoremname}
\newenvironment{namedthm}[1]
  {\renewcommand{\thistheoremname}{#1}%
   \begin{genericthm}}
  {\end{genericthm}}
  
\crefname{table}{Protocol}{Protocols}

\newcommand\blfootnote[1]{%
  \begingroup
  \renewcommand\thefootnote{}\footnote{#1}%
  \addtocounter{footnote}{-1}%
  \endgroup
}

\begin{document}
\floatstyle{boxed}
\newfloat{protocol}{htb}{lop}
\floatname{protocol}{Protocol }
\setlength{\parskip}{1pt}

\title{ Loss-tolerant quantum secure positioning with weak laser sources }

\author{Charles Ci Wen Lim}\email{limc@ornl.gov}
\affiliation{Quantum Information Science Group, Computational Sciences and Engineering Division,
Oak Ridge National Laboratory, Oak Ridge, Tennessee 37831-6418, USA}
\author{Feihu Xu}
\affiliation{Research Laboratory of Electronics, Massachusetts Institute of Technology, 77 Massachusetts Avenue, Cambridge, Massachusetts 02139, USA}
\author{George Siopsis}
\affiliation{
 Department of Physics and Astronomy, The
University of Tennessee, Knoxville, Tennessee 37996-1200, USA}
\author{Eric Chitambar}
\affiliation{Department of Physics and Astronomy, Southern Illinois University, Carbondale, Illinois 62901, USA}
\author{Philip G. Evans}
\affiliation{Quantum Information Science Group, Computational Sciences and Engineering Division,
Oak Ridge National Laboratory, Oak Ridge, Tennessee 37831-6418, USA}
\author{Bing Qi}
\affiliation{Quantum Information Science Group, Computational Sciences and Engineering Division,
Oak Ridge National Laboratory, Oak Ridge, Tennessee 37831-6418, USA}
\affiliation{
 Department of Physics and Astronomy, The
University of Tennessee, Knoxville, Tennessee 37996-1200, USA}

\begin{abstract}
Quantum position verification (QPV) is the art of verifying the geographical location of an untrusted party.~Recently, it has been shown that the widely studied Bennett \& Brassard 1984 (BB84) QPV protocol is insecure after the 3 dB loss point assuming local operations and classical communication (LOCC) adversaries.~Here, we propose a time-reversed entanglement swapping QPV protocol (based on measurement-device-independent quantum cryptography) that is highly robust against quantum channel loss.~First, assuming ideal qubit sources, we show that the protocol is secure against LOCC adversaries for any quantum channel loss, thereby overcoming the 3 dB loss limit.~Then, we analyze the security of the protocol in a more practical setting involving weak laser sources and linear optics.~In this setting, we find that the security only degrades by an additive constant and the protocol is able to verify positions up to 47 dB channel loss.
\end{abstract}

\maketitle

\section{Introduction}\blfootnote{This manuscript has been authored by UT-Battelle, LLC under Contract No.~DE-AC05-00OR22725 with the U.S. Department of Energy.~The United States Government retains and the publisher, by accepting the article for publication, acknowledges that the United States Government retains a non-exclusive, paid-up, irrevocable, worldwide license to publish or reproduce the published form of this manuscript, or allow others to do so, for United States Government purposes.~The Department of Energy will provide public access to these results of federally sponsored research in accordance with the DOE Public Access Plan (http://energy.gov/downloads/doe-public-access-plan).}
How can one verify that an untrusted party (someone with no credentials) is indeed at a particular geographical location?~In cryptography, this problem is closely related to the task of \emph{position verification}, where a prover $P$ has to convince a set of remote verifiers  $V_1,V_2,\ldots,$ that he or she is at a certain geographic position $\mathsf{pos}^*$~\cite{Chandran09}.~At the end of the task, the verifiers either agree or disagree with the prover:~agreement means the prover gains a geographical credential, while disagreement means the prover remains with zero credentials.~Beyond position verification, such geographical credentials can also be used to build other cryptographic tasks like authentication and key distribution~\cite{Buhrman14}.

In the classical setting, it has been shown that position verification is insecure against unbounded adversaries~\cite{Chandran09}.~This impasse is mainly due to the fact that colluding adversaries can retrieve, store, and share classical challenges with each other.~One solution is to adopt the so-called \emph{bounded-retrieval model} and limit the amount of information that an adversary can retrieve from the public channel~\cite{Chandran09}.~However, this model is difficult to justify in practice.~Drawing insights from the bounded-retrieval model, researchers proposed quantum position verification (QPV) as a means to achieve information-theoretic security~\cite{Kent06, Kent11, Kent112, Malaney10, Lau11, Buhrman14}.~The basic idea is to replace classical challenges with quantum challenges (quantum states) and utilize the \emph{quantum no-cloning principle} to bound the amount of retrievable information.~Unfortunately, this intuition is not enough to guarantee unconditional security in the quantum setting, as colluding adversaries can make use of preshared entanglement to perform nonlocal computation with one round of classical communication~\cite{Vaidman03, Clark2010, Lau11, Beigi11, Buhrman14}.~In light of these impossibility results, the most obvious solution is to consider adversaries with no preshared entanglement, a scenario that is known as the NPE-model~\cite{Buhrman14}.~Assuming perfect channel transmittance, the Bennett \& Brassard 1984 (BB84) QPV protocol has been proven secure against the NPE-model~\cite{Buhrman14}, and more generally against adversaries with linearly bounded entanglement~\cite{Beigi11,Tomamichel13, Ribeiro15, Chakraborty15}. 

In the case of high quantum channel loss, it turns out that the situation is much more constrained. In particular, it has been shown that BB84 QPV is highly vulnerable against loss-dependent attacks and is insecure after the 3 db loss point~\cite{Qi15}.~This weakness is in part due to the design of the verification challenge.~To see this, recall that in BB84 QPV, one verifier $V_1$ sends a qubit prepared in one of the four BB84 states to the prover $P$, while the other verifier $V_2$ sends the basis information. Then, the prover is asked to extract the encoded bit from the qubit using the received basis information.~Now, if the quantum channel loss is sufficiently high, then the adversaries can break the protocol with the following local operations and classical communication (LOCC) attack.~First, the adversary nearest to $V_1$ (called $E_1$) measures $V_1$'s qubit in a randomly chosen basis and sends the measurement result and the basis choice to the other adversary $E_2$, who is located next to $V_2$.~Likewise, $E_2$ duplicates the basis information of $V_2$ and sends a copy to  $E_1$.~Finally, the adversaries report $E_1$'s measurement outcome to their respective verifiers if the basis choices of $E_1$ and $V_2$ are the same.~Otherwise, they claim no detection.~Evidently, this attack works whenever the quantum channel loss is greater than 1/2, thus implying a 3 dB loss limit.~More crucially, this means that BB84 QPV is not useful in practice as most free space quantum communication systems have more than 3 dB loss~\cite{Aspelmeyer2003}.

One way to overcome the above limitation is to go beyond the BB84 encoding scheme and encode the qubits in more than two bases.~More concretely, if the number of possible encoding bases is $N$, then the above LOCC attack can only succeed with probability $1/N$.~Following this intuition, it has been shown that multi-basis QPV using weak laser sources is secure against specific LOCC attacks up to $13$ dB loss and 0.01 quantum bit error rate~\cite{Qi15}.~Another solution is to use quantum memories and separate the quantum transmission phase from the (classical) basis distribution phase~\cite{Malaney10}.~That is, the quantum challenge (a collection of quantum states) is first delivered to the prover and stored in a quantum memory.~Then, the verifiers only send the classical challenge after the prover confirms that the quantum challenge has been received.~Thus assuming perfect classical communication, the protocol is essentially secure against loss-dependent attacks.~However, such protocol may require long-lived quantum memories.
\begin{figure}[t]
  \includegraphics[width=85mm]{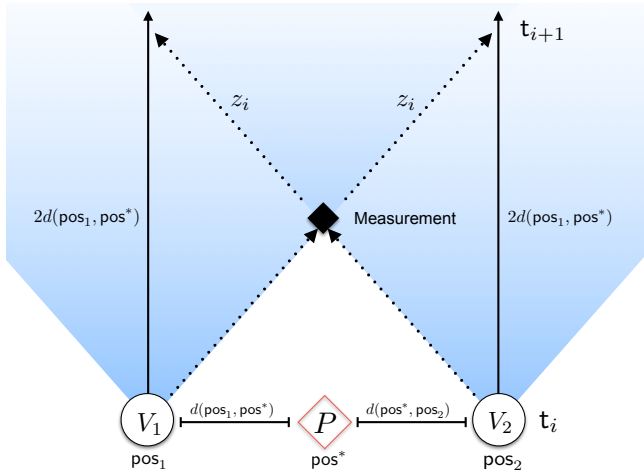}
  \caption{\textbf{Relativistic constraints.}~We assume that all quantum and classical signals travel at the speed of light and that the speed of light is normalized to unity.~In this case, the time required to send a message from one position to another position is equal to the Euclidean distance between them. More specifically, the Euclidean distance between $\mathsf{pos}_1$ and $\mathsf{pos}^*$ is defined as $d(\mathsf{pos_1},\mathsf{pos}^*)$ where $d(\cdot,\cdot)$ is the distance measure in $\reals$.~The protocol is based on a $N$-fold sequential repetition setting, where the verifiers only send out their qubit states at intervals of $\mathsf{t}_{i+1}-\mathsf{t}_{i}=2d(\mathsf{pos_1},\mathsf{pos}^*)=2d(\mathsf{pos}^*,\mathsf{pos}_2)$.~Note that for simplicity we assume the prover is located at the center.    }\label{fig1:spacetimecone}
\end{figure}

Here, we present a QPV protocol that is secure against LOCC adversaries for any quantum channel loss.~The protocol is based on the concept of measurement-device-independent quantum key distribution (MDI-QKD)~\cite{Lo12} and uses time-reversed entanglement swapping to check for quantum correlations~\cite{Qi15b}.~The basic idea is that if the prover is indeed at the claimed position, then he or she should be able to perform a \emph{local} entangling measurement on the verifiers' BB84 qubits and create quantum correlations between them (as in entanglement swapping). However, if the prover is dishonest and is not at the claimed position, then by definition he or she can only collude with other dishonest provers to perform LOCC measurements on the qubits.~In this case, no quantum correlations can be created between the verifiers.~Therefore, by comparing the measured error rate against some tolerated error rate, the verifiers can check if the prover is at the claimed position or not.~Furthermore, like MDI-QKD, our QPV protocol does not require quantum memories and can be implemented with weak laser sources, linear optics and standard single photon detectors.

For practical reasons, we consider the sequential multi-round setting where the verifiers only send out their BB84 qubits after receiving the measurement outcome from the previous round.~In this setting, the standard relativistic constraints (see Fig.~\eqref{fig1:spacetimecone}) only apply to each individual round.~One of the main advantages of sequential multi-round is that the adversaries are limited to independent attacks (also known as \emph{collective attacks} in quantum cryptography), which greatly simplifies the security analysis.~However, sequential multi-round setting includes the possibility that the adversaries could use the first round to distribute entanglement for later rounds and break the protocol.~To rule out such a possibility, the most consistent solution, arguably, is to assume LOCC adversaries, which by definition precludes the distribution of entanglement at any point in the protocol.~Alternatively, we can also keep the NPE-model and further assume the adversaries lose their entanglement at the start of every round.~In this work, we consider security against LOCC adversaries and leave the security of NPE-model for future work.~Here, it is implicit that security against LOCC adversaries means security against LOCC attacks that are compatitible with the underlying relativistic constraints (i.e., those with one round of classical communication).

The paper is organized as follows.~For pedagogical reasons, in Section \ref{sec:2} we first present the details of our QPV protocol with ideal BB84 qubit states (called \cref{protocol:qubit}).~Then, in Section \ref{sec:3} we analyze the security of our qubit protocol against LOCC adversaries.~In Section \ref{sec:4}, we extend \cref{protocol:qubit} to weak laser sources based on the decoy-state method~\cite{Decoy} (called \cref{protocol:decoy}) and derive its security bound.~Finally in Section \ref{sec:dis}, we conclude with a discussion on possible future work.

\section{Qubit Protocol}
\label{sec:2}
For simplicity, we consider the one dimensional scenario where everyone is positioned on a straight line.~In this scenario, the verifiers are assumed to have access to a private classical channel~\cite{footnote1} and each verifier is equipped with a local source of randomness and a trusted BB84 qubit preparation device.~More specifically, each qubit preparation device accepts two bits $k_1,k_2$ as an input and generates $\omega_{k_1,k_2}$ using
\begin{eqnarray*}
\omega_{0,0}&:=&\frac{\mathbb{I}+\mathbb{X}}{2},\quad\omega_{0,1}:=\frac{\mathbb{I}-\mathbb{X}}{2},\\
\omega_{1,0}&:=&\frac{\mathbb{I}+\mathbb{Y}}{2},\quad\omega_{1,1}:=\frac{\mathbb{I}-\mathbb{Y}}{2},
\end{eqnarray*}
where $\mathbb{X}$ and $\mathbb{Y}$ (together with $\mathbb{Z})$ are the standard Pauli matrices.~Our QPV protocol is framed in a $m$-fold sequential repetition picture and is characterized by two threshold parameters, i.e., the tolerated number of detection events, $n_\tn{th}$, and the tolerated error rate, $\delta_\tn{th}< 1/4$.~The protocol concludes by outputting either $\{\mathtt{Y}, \mathtt{N}\}$, where $\mathtt{Y}$ means agreement and $\mathtt{N}$ means disagreement.~Below, we describe our protocol in more detail.  

\begin{table}[h!]
\hrule 
\smallskip
\textbf{{Protocol with ideal BB84 qubits}}  
\smallskip
\hrule
\justify
\noindent \textbf{1.~Preparation.}~The preparation phase is carried out $i=1,2,\ldots,m$ times, one after the other.~In each $i$th run, the verifiers first use the private classical channel to generate a random basis choice $b_i$.~Then, they each generate a random bit (which we denote by $x_i$ and $y_i$, respectively) and use it to prepare a qubit and send it to the prover.~The transmission is synchronized in such a way that the qubits reach $\mathsf{pos}^*$ at time $\mathsf{t}_i+\tau$, where $\tau=d(\mathsf{pos_1},\mathsf{pos}^*)=d(\mathsf{pos}^*,\mathsf{pos}_2)$, i.e., see Fig.~\eqref{fig1:spacetimecone}. \\

\noindent \textbf{2.~Measurement.}~The prover makes an entangling measurement on $\omega_{b_i,x_i}\otimes \omega'_{b_i,y_i}$ and obtains one of the three possible outcomes, $z_i \in \{0, 1,\varnothing\}$.~The outcome is then reported to the verifiers.\\

\noindent \textbf{3.~Quota check.}~The verifiers accept the measurement outcome $z_i$ only if it arrives in time.~If one of the outcomes does not arrive in time or the verifiers receive different outcomes, they abort the protocol and output $\mathtt{N}$.~If the protocol does not abort at the end of the measurement phase,~the verifiers perform a \emph{quota check}:~they calculate $s_{1,1}=|\CMcal{Z}|$, where $\CMcal{Z}=\{i:z_i \not = \varnothing\}$, and check if $s_{1,1}\geq n_\tn{th}$.  If the check is positive, they select a random subset $\CMcal{Z}'$ of size $n_\tn{th}$ from $\CMcal{Z}$. Otherwise, they abort and output $\mathtt{N}$.\\ 

\noindent \textbf{4.~Verification.}~Conditioned on passing the quota check, the verifiers compute the error rate and check if\end{table}
\begin{table} \justify 
\[
\hat{\delta}_\tn{test}=\frac{r_{1,1}}{s_{1,1}} \leq \delta_\tn{th},
\] where~$r_{1,1}=|\CMcal{E}|$ and $\CMcal{E}=\{i:z_i \not = x_i \oplus y_i | z_i \in \CMcal{Z}'\}$.~If the check is positive, they agree with the prover and output $\mathtt{Y}$, otherwise they output $\mathtt{N}$.\\

\hrule
  \caption{QPV with BB84 qubits. 
    \label{protocol:qubit}}
\end{table}

\begin{figure}[t!]
\centerline{\includegraphics[width=7.8cm]{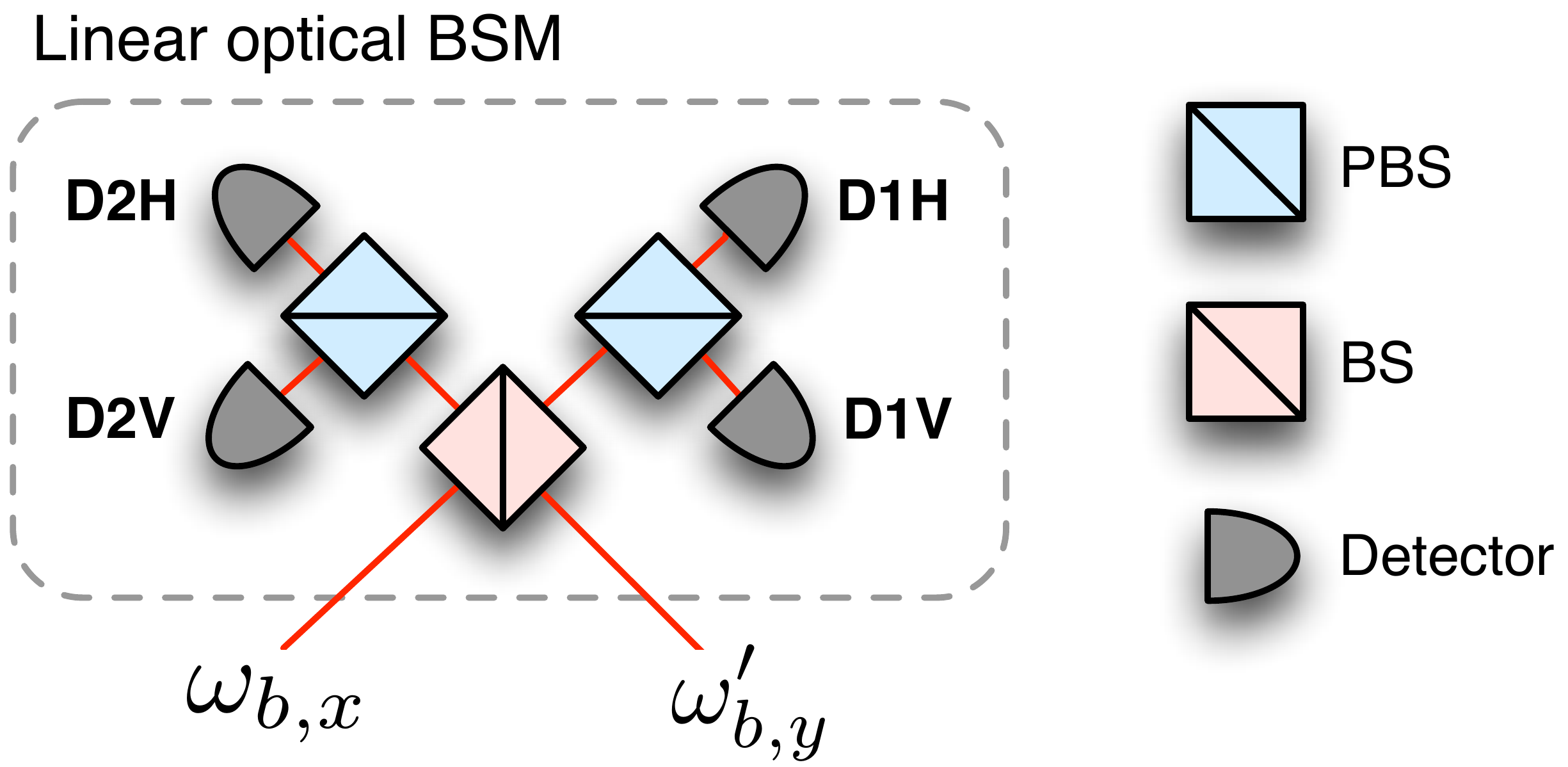}} \caption{\textbf{BSM based on linear optics.}~A successful Bell state measurement corresponds to the following detection patterns: a coincident detection in $D_{1H}$ and $D_{2V}$, or in $D_{1V}$ and $D_{2H}$, indicates a projection into the Bell state $|\Psi^{-}\rangle$, while a click in $D_{1H}$ and $D_{1V}$ , or in $D_{2H}$ and $D_{2V}$ , reveals a projection into the Bell state $|\Psi^{+}\rangle$}\label{fig2:BSM_LO}
\end{figure}

Let us first present an optical implementation based on single-photon sources and linear optics which shows that the above protocol is cryptographically complete (see Section \ref{sec:3} for a brief discussion and Ref.~\cite{Chandran09} for a more formal definition).~Starting from the preparation phase, the verifiers each use their randomly generated bit values ($k_1, k_2$) to prepare one of the four possible polarized single-photon states, $\{(\ket{H}+(i)^{k_1}(-1)^{k_2}\ket{V})/\sqrt{2}\}$, and send it to the prover.~Assuming linear optics, the prover can implement a Bell-state measurement (BSM) with $1/2$ efficiency, i.e., one that is capable of discriminating between two Bell states~\cite{Vaidman1999, Lutkenhaus1999}~(see Fig.~(\ref{fig2:BSM_LO})).~In this case, the expected error rate and detection rate are 0 and $1/2$, respectively.~That is, whenever the verifiers send the same polarized state (i.e., $x_i=y_i$), they get $\Psi^+$ (i.e., $z=0$) with probability $1/2$, $\Psi^-$ (i.e., $z=1$) with zero probability, and an inconclusive outcome with probability $1/2$.  For different polarized states (i.e., $x_i \not= y_i$), they get $\Psi^+$ with zero probability, $\Psi^-$ with probability $1/2$, and an inconclusive outcome with probability $1/2$.~Therefore, the verifiers will always agree with the honest prover if $n_\tn{th} \leq m/2$ is chosen. In this case, the protocol is perfectly complete in the asymptotic limit.

\section{Security of qubit protocol}
\label{sec:3}

From a prepare \& measure perspective, the basic idea of our protocol is to have the prover guess the XOR of the verifiers' bit values.~That is, in each round of the protocol the prover is given a random joint state $\omega_{b,x} \otimes \omega'_{b,y}$ and is supposed to guess the underlying $x\oplus y$. The main security principle of  \cref{protocol:qubit} is that the best measurement (i.e., one that gives the highest guessing probability) is necessarily an entangling measurement, which according to our security model is only possible at the claimed position $\mathsf{pos}^*$. As we will soon see below, LOCC adversaries (due to their limited measurement possibilities) can only guess $x\oplus y$ with at most probability 3/4.

To start with, the most general strategy is to maximize the guessing probability over all two-qubit positive-operator valued measure (POVM) operators $\{\Pi_z\}_{z={0,1,\varnothing}}$ constrained to an average quantum channel loss parameter (denoted by $\eta$). Mathematically, the maximum guessing probability is given by
\be \label{eq:guessing}
P^\tn{max}_\tn{guess}(\eta):=\max_{\{\Pi_z\}_z}\frac{1}{2}\frac{\Tr\left[ \rho_0 \Pi_0+\rho_1\Pi_1\right]}{\eta},
\ee
where 
\[
\rho_{0}:=\frac{1}{4}\!\!\!\!\!\!\!\sum_{\substack{b,x,y\\ \tn{s.t.}\,x\oplus y=0}}\!\!\!\!\!\!\!\omega_{b,x}\otimes \omega'_{b,y},\quad \, \rho_{1}:=\frac{1}{4}\!\!\!\!\!\!\!\sum_{\substack{b,x,y\\ \tn{s.t.}\,x\oplus y=1}}\!\!\!\!\!\!\!\omega_{b,x}\otimes \omega'_{b,y},\] 
and $\Tr[\rho_i \Pi_\varnothing]=1-\eta$ for $i=0,1$.~Note that for $\eta=1$, Eq.~\eqref{eq:guessing} is given by the Helstrom's bound~\cite{Helstrom1969}, i.e., $P^{\tn{max}}_\tn{guess}(1)=1/2+ \|\rho_0-\rho_1\|_1/4=3/4$.

In the case of dishonest LOCC prover(s), the maximum guessing probability is 
\be \label{eq:guessing_locc}
P^\tn{max}_\tn{guess}(\eta|\tn{LOCC}):=\max_{\{\Pi^{\tn{LOCC}}_z\}_z}\frac{1}{2}\frac{\Tr\left[ \rho_0 \Pi_0+\rho_1\Pi_1\right]}{\eta},
\ee
where the maximization is now taken over all two-qubit LOCC measurements.~This maximization problem is however difficult to solve as the mathematical characterization of LOCC measurements is highly complex (even for two-qubit measurements with one round of communication).~To overcome this problem, we use a circuitous approach based on positive partial transpose (PPT) measurements which admit two advantages over LOCC measurements.~First, the set of LOCC measurements is a proper subset of PPT  measurements, which means the guessing probability taken over all PPT measurements is necessarily an upper bound  on Eq.~\eqref{eq:guessing_locc}, i.e., $P^{\tn{max}}_\tn{guess}(\eta|\tn{PTT}) \geq P^{\tn{max}}_\tn{guess}(\eta|\tn{LOCC})$. Second, we may reformulate the maximization of $P^{\tn{max}}_\tn{guess}(\eta|\tn{PTT})$ as a semidefinite program (SDP)~\cite{SDP1996}, where the optimization is taken over all two-qubit positive operators satisfying the PPT condition (which in turn is represented by a set of linear and positive semidefinite conditions)~\cite{Cosentino2013,Lim2016}. More concretely, we may express the maximization of $\eta P^{\tn{max}}_\tn{guess}(\eta|\tn{PTT})$ (for a fixed $\eta$) as \begin{eqnarray*}
{\tt{maximize}}&:&\frac{1}{2}\Tr\left[ \rho_0 \Pi_0+\rho_1\Pi_1\right]\\ 
{\tt{subject~to}}&:&  \Pi_0+ \Pi_1+ \Pi_{\varnothing} = \mathds{1},\\  
&& \Tr[\rho_i \Pi_{\varnothing}]=1-\eta,\quad i=0,1\\
&& \Pi_k^{T_\mathsf{B}} \succeq 0,\quad k=0,1,\varnothing,
\end{eqnarray*}
where $T_\mathsf{B}$ means the partial transpose with respect to the measurement on the second qubit. The optimal solution to the above SDP (primal program) is $3/4\eta$ (see Appendix \ref{App:ss1}), which implies the guessing probability for LOCC adversaries is upper bounded by\be \label{eq:separablebound}
P^{\tn{max}}_\tn{guess}(\eta|\tn{LOCC}) \leq \frac{3}{4}.
\ee
Interestingly, we see that $P^{\tn{max}}_\tn{guess}(\eta|\tn{LOCC})$ is bounded by a constant term that is independent of the detection efficiency $\eta$.~In fact, it can be shown that this bound is tight, i.e., there exists a LOCC measurement that reaches the PPT bound for any $\eta$.~To show this, suppose that there are two adversaries, $E_1$ and $E_2$, who are positioned next to $V_1$ and $V_2$, respectively.~Furthermore, suppose that they share a source of shared randomness, $\lambda$, which takes value from $\{0,1\}$ with probabilities $\Pr[\lambda=0]=1-\eta$ and $\Pr[\lambda=1]=\eta$, respectively.~Now, in each round of the protocol, if $\lambda=1$, the adversaries measure their respective qubits in the diagonal basis $\mathbb{X}$ and exchange the measurement outcomes.~Then, they compute the XOR of their outcomes and send it to the verifiers.~If $\lambda=0$, they jointly report no detection.~Using this measurement strategy, it can be easily verified that the guessing probability is 3/4 for any detection efficiency.~Alternatively, the upper bound can also be reached by using the $\mathbb{Y}$ basis, or using a statistical mixture of $\mathbb{X}$ and $\mathbb{Y}$ bases with the aid of additional shared randomness.~

From the above, it is clear that no coalition of LOCC adversaries can correctly predict $x\oplus y$ even if $\eta$ is arbitrarily small.~Coupled with the earlier example that an honest prover (who is at the claimed position and using linear optics) is able to correctly predict $x\oplus y$ for $\eta \leq 1/2$, it follows that a conclusive verification of the prover's geographical position is equivalent to checking if the expected error rate is smaller than the minimum LOCC error rate, $\delta_\tn{LOCC}:=1-P^{\tn{max}}_\tn{guess}(\eta|\tn{LOCC})=1/4$. 
 
Before we present the security of \cref{protocol:qubit}, let us first briefly explain and define what it means for the protocol to be secure.~The security of a generic QPV protocol is generally analyzed using two conditions, namely the completeness condition and the soundness condition~\cite{Chandran09}. The completeness condition, roughly speaking, is a measure of how often the protocol will agree with an honest prover.~Note that in the preceding section, we have already shown (using an ideal optical model) that \cref{protocol:qubit} is perfectly complete in the asymptotic limit for $n_\tn{th} \leq m/2$.~The soundness condition, which we will be analyzing in more detail below, is a conservative measure of how often the protocol will agree with a coalition of adversaries.~More precisely, the soundness condition (adapted to our security model) is defined as

\begin{definition}[Soundness]\emph{The protocol is said to be $\varepsilon$-sound if for any coalition of \tn{LOCC} adversaries ${E}_1, {E}_2, {E}_3,\ldots,$ at positions $\mathsf{pos}'_1,\mathsf{pos}'_2,\mathsf{pos}'_3,\ldots \not= \mathsf{pos}^*$ and using resources only at
these positions, the verifiers agree with probability at most $\varepsilon$.
 }\end{definition}
 
The goal of the security analysis is to compute an upper bound on the soundness error, $\varepsilon$, in terms of the protocol parameters, i.e., the tolerated number of detection events, $n_\tn{th}$, and the tolerated error rate, $\delta_\tn{th}$.

\begin{namedthm}{Result}[Security with qubits]\emph{Given $n_\tn{th}$ and $\delta_\tn{th}$, the protocol is $\varepsilon_\tn{qubit}$-sound against \tn{LOCC} adversaries with}
\be\label{eq:soundness_qubit}
\varepsilon_\tn{qubit} \leq e^{ -2n_\tn{th}\left(1/4-\delta_{\tn{th}}\right)^2 }.
\ee
\end{namedthm}

\begin{proof}The soundness of the protocol is obtained by asking what is the maximum probability that the verifiers agree with the adversaries.~In what follows, for brevity reasons, we will denote the event that the protocol passes the quota check by $\Omega_\tn{qc}$, and omit the conditioning on LOCC attacks (since this is clear in the context).~First, we note that the soundness error is upper bounded by the probability that the verifiers agree with adversaries conditioned on $\Omega_\tn{qc}$, i.e.,
\begin{eqnarray*} 
\varepsilon_\tn{qubit}&=&\Pr[\Omega_\tn{qc}] \Pr[\mathtt{Y}|\Omega_\tn{qc}]+\Pr[\Omega_\tn{qc}^\tn{c}] \Pr[\mathtt{Y}|\Omega_\tn{qc}^\tn{c}]\\&\leq&\Pr[\mathtt{Y}|\Omega_\tn{qc}],
\end{eqnarray*} where we used $\Pr[\Omega_\tn{qc}]  \leq 1$ and $\Pr[\mathtt{Y}|\Omega_\tn{qc}^\tn{c}]=0$ to get the inequality.~Next, we note that the protocol outputs $\mathtt{Y}$ only if the measured error rate $\hat{\delta}_\tn{test}$ is less than or equal to the tolerated error rate $\delta_\tn{th}$.~This gives
\[ \varepsilon_\tn{qubit}  \leq \Pr[\mathtt{Y}|\Omega_\tn{qc}]= \Pr[\hat{\delta}_\tn{test} \leq \delta_\tn{th}|\Omega_\tn{qc}]. \]
The above probability term can be modeled by a Bernoulli experiment with $n_\tn{th}$ trials.~More precisely, for each element in $\CMcal{Z}'$, let $\rv{W}_i$ be an indicator random variable taking values in $\{0,1\}$, where $0$ means no error and $1$ means otherwise.~Let $\hat{\delta}_\tn{test}=\sum_{i=1}^{n_\tn{th}}\hat{W}_i/n_\tn{th}$, then the probability of $\tn{E}[\hat{\delta}_\tn{test}]-\hat{\delta}_\tn{test} \geq \beta$ for some $\beta>0$ is bounded by the Hoeffding's inequality~\cite{hoeffding63}:
\[
\Pr[\tn{E}[\hat{\delta}_\tn{test}]-\hat{\delta}_\tn{test} \geq \beta ] \leq e^{ -2n_\tn{th}\beta^2}.
\] Finally, by setting $\tn{E}[\hat{\delta}_\tn{test}]=\delta_\tn{LOCC}$, and $\beta=1/4-\delta_\tn{th}$, we have
\[
\varepsilon_\tn{qubit} \leq \Pr[ {\delta}_\tn{th}\geq \hat{\delta}_\tn{test} |\Omega_\tn{qc}] \leq e^{-2n_\tn{th}\left(1/4-\delta_\tn{th}\right)^2}.
\] 
\end{proof}From the above, we see that the soundness error is exponentially small in $n_\tn{th}\left(1/4-\delta_\tn{th}\right)$. This means that \cref{protocol:qubit} can be made highly reliable by choosing a large $n_\tn{th}$ and a stringent error threshold (i.e., a small $\delta_\tn{th}$).~More importantly, the soundness error is independent of the detection rate, which means that \cref{protocol:qubit} is secure against arbitrary quantum channel loss.

\section{Decoy-state Method}
\label{sec:4}
In \cref{protocol:qubit} we have assumed that the verifiers are able to reliably prepare ideal qubit states.~However in practice, this assumption is unrealistic as it requires true single-photon sources.~A more practical option is to use weak laser sources, which are good approximations of probabilistic single-photon sources.~More concretely, the output of a laser with intensity $\mu=|\alpha|^2$ is described by a coherent state, $\ket{\alpha}=e^{-\mu/2}\sum_{n=0}\alpha^n/\sqrt{n!}\ket{n}$, where $\{\ket{n}\}_n$ is the photon number (Fock) basis.~Assuming that the laser is phase randomized, the photon number of each output state follows a Poisson distribution with its mean given by the laser's intensity~\cite{phaserandomize}.~In this case, the output state is described by\begin{equation} \label{Model:CoherentState}
\begin{aligned}\nonumber
\rho_\tn{laser} = \frac{1}{2\pi} \int\limits_{0}^{2\pi} d \theta \ket{ |\alpha| e^{i\theta} } \bra{ |\alpha| e^{i\theta} }
= \sum^{\infty}_{n=0}\frac{\mu^n}{n!}e^{-\mu} \ket{n}\bra{n},
\end{aligned}
\end{equation}
where $\theta$ is the phase of the state and $\ket{n}\bra{n}$ is the density matrix of the $n$-photon state.~This means that in each round, the laser source emits a vacuum state with probability $e^{-\mu}$, a single photon state with probability $\mu e^{-\mu}$, and a multi-photon state with probability $1-(1+\mu)e^{-\mu}$.~Thus, we may think of weak laser sources as probabilistic single-photon sources if the laser intensity is sufficiently small.

However, in the case of QPV, the non-vanishing multi-photon probability is a major security issue, especially when the quantum channel loss is high.~In particular, colluding adversaries can postselect on laser pulses with 3 photons or more and perform unambiguous state discrimination to determine the verifier's basis and bit information with success probability $\geq$~1/2~\cite{Scarani2004}.~If the quantum channel loss is high enough, then it is not hard to see that QPV is reduced to the classical version (with classical challenges) when all $n<3$ laser pulses are blocked and returned as empty detections.~Importantly, this implies that the security of QPV with weak laser sources is not independent of the quantum channel loss.

In the following, we will show that QPV with weak laser sources is still highly robust against quantum channel loss, tolerating up to 47 dB loss assuming realistic parameters.~The central idea is to use the decoy state method~\cite{Decoy} to estimate the number of single-photon detections, i.e., the number of instances in which both verifiers send single-photon states and a successful BSM outcome is announced (denoted by $s_{1,1}$), and the number of errors in these single-photon detections (denoted by $r_{1,1}$)~\cite{Xu13, Curty14}.~Then by using these estimates, the verifiers can verify the position of the prover by checking if the estimated single-photon error rate is smaller than the tolerated error rate (as in \cref{protocol:qubit}).

We consider a decoy-state method with three intensities, $\CMcal{I}:=\{\mu_1, \mu_2, \mu_3\}$, where $\mu_1 > \mu_2 + \mu_3$ and $\mu_2> \mu_3 \geq 0$.~The relevant estimates are (1) a lower bound on $s_{1,1}$ and (2) an upper bound on $r_{1,1}$, which we denote by random variables $\rv{s}^\tn{lb}_{1,1}$ and $\rv{r}^\tn{ub}_{1,1}$, respectively.~Accordingly, this means that there are two possible statistical errors, one due to the estimation of ${s}_{1,1}$ and the other due to the estimation of ${r}_{1,1}$.~The reliability of these estimates is parameterized by a non-negative security parameter, $\nu$.~Below we present the protocol in more detail. 
\begin{table}
\hrule 
\smallskip
\textbf{{Protocol with decoy-state method}}  
\smallskip
\hrule
\justify
\noindent\textbf{1.~Preparation.}~The prepare \& measurement phase is carried out $i=1,2,\ldots,m$ times, one after the other.~Like in the qubit protocol, the verifiers agree on a random basis choice $b_i$ using the private classical channel, and they each independently generate a random bit.~For the decoy-state method, they each select an intensity value from $\CMcal{I}:=\{\mu_1, \mu_2, \mu_3\}$ with probabilities $p_{\mu_1}$, $p_{\mu_2}$, and $p_{\mu_3}$, respectively.~We write $g_i$ and $h_i$ to denote their respective intensity choices for each $i$th round.~Finally, the verifiers each prepare a weak laser pulse based on their generated values and send the encoded laser pulse to the prover. \\

\noindent \textbf{2.~Measurement.}~The prover makes an entangling measurement on the laser pulses and report the outcome, $z_i \in \{0, 1,\varnothing\}$, back to the verifiers. \\

\noindent \textbf{3.~Quota check.}~Similar to the qubit protocol, the verifiers only accept the measurement outcomes if they are consistent with the timing constraints.~If one of the outcomes does not meet the timing constraint or the verifiers receive different outcomes, the protocol aborts and the verifiers output $\mathtt{N}$.~If the protocol does not abort at the end of the measurement phase, the verifiers perform a quota check.~Setting $n_\tn{obs}^{u, v}=|\CMcal{Z}^{u, v}| $ for $u, v=\mu_1, \mu_2, \mu_3$ and  $n_{\tn{obs}}=\sum_{u,v}n_\tn{obs}^{u, v}$, the verifiers compute a lower bound on $s_{1,1}$ (see Appendix \ref{App:ss2}) using 
\be \label{Eq:decoy_1}
\hat{s}_{1,1}^{\tn{lb}}\!=\!\left\lfloor\!\frac{(\mu_\rs^2\!-\!\mu_\rdd^2)(\mu_\rs\!-\!\mu_\rdd)\gamma_2 
- (\mu_\rd^2\!-\!\mu_\rdd^2)(\mu_\rd\!-\!\mu_\rdd)\gamma_1}{(\mu_\rs -\mu_\rdd)^2(\mu_\rd-\mu_\rdd)^2(\mu_\rs-\mu_\rd)}\!\right\rfloor \!,
\ee where
\begin{multline*}
\gamma_1:=\chi^{\mu_\rs,\mu_\rs}+\chi^{\mu_\rdd,\mu_\rdd}-\chi^{\mu_\rs,\mu_\rdd}-\chi^{\mu_\rdd,\mu_\rs}\\ + \nu^{\frac{1}{2}} n_\tn{obs}^{\frac{1}{2}}\left(\xi^{\mu_\rs,\mu_\rs}+\xi^{\mu_\rdd,\mu_\rdd}+2\xi^{\mu_\rs,\mu_\rdd}\right),
\end{multline*} 
\begin{multline*}
\gamma_2:=\chi^{\mu_\rd,\mu_\rd}+\chi^{\mu_\rdd,\mu_\rdd}-\chi^{\mu_\rd,\mu_\rdd}-\chi^{\mu_\rdd,\mu_\rd}\\ - \nu^{\frac{1}{2}} n_\tn{obs}^{\frac{1}{2}}\left(\xi^{\mu_\rd,\mu_\rd}+\xi^{\mu_\rdd,\mu_\rdd}+2\xi^{\mu_\rd,\mu_\rdd}\right),
 \end{multline*} 
with $\xi^{u,v}:=\exp(u+v)p_u^{-1}p_v^{-1}$ and $\chi^{u,v}:=\xi^{u,v}n_\tn{obs}^{u,v}$ for all $u,v\in \CMcal{I}$.~The verifiers proceed to the verification step if \[s_{1,1}^\tn{lb}\geq n_{\tn{th}},\] otherwise they abort the protocol and output $\mathtt{N}$. \\

\noindent \textbf{5.~Verification.}~The verifiers first calculate the number of errors (denoted by $m_{\tn{obs}}^{u,v}$) in each $\CMcal{Z}^{u,v}$ and the total number of errors, $m_\tn{obs}=\sum_{u,v}m^{u,v}_{\tn{obs}}$.~Then, they compute an upper bound on $ r_{1,1}$ using
\be \label{Eq:decoy_2}
\rv{r}_{1,1}^{\tn{ub}}=\min \left\{ \left \lceil \frac{\gamma_3}{(\mu_\rd-\mu_\rdd)^2} \right\rceil, \left \lceil \frac{\rv{s}_{1,1}^\tn{lb}}{2} \right\rceil\right\}
\ee
where
\begin{multline*}\gamma_3:= \zeta^{\mu_\rd,\mu_\rd}+\zeta^{\mu_\rdd,\mu_\rdd}-\zeta^{\mu_\rd,\mu_\rdd}-\zeta^{\mu_\rdd,\mu_\rd} \\ + \nu^{\frac{1}{2}} m_\tn{obs}^{\frac{1}{2}}\left(\xi^{\mu_\rd,\mu_\rd}+\xi^{\mu_\rdd,\mu_\rdd}+2\xi^{\mu_\rd,\mu_\rdd}\right).
\end{multline*} with $\zeta^{u,v}:=\xi^{u,v}m_\tn{obs}^{u,v}$.~Finally, the verifiers agree with the prover and output $\mathtt{Y}$ if \end{table}
\begin{table} \justify 
\[
\hat{\delta}_\tn{test}^{\tn{decoy}}=\frac{\rv{r}_{1,1}^\tn{ub}}{\hat{s}_{1,1}^\tn{lb}} \leq\delta_{\tn{th}},
\]
Otherwise, they output $\mathtt{N}$. \\ 
\hrule
  \caption{QPV with decoy state method. 
    \label{protocol:decoy}}
\end{table}
\section{Security analysis and simulation}
A crucial step in the above security analysis of \cref{protocol:qubit} is that the verifiers are able to directly observe $s_{1,1}$ and $r_{1,1}$ and check if the protocol has sufficient statistics, i.e., $s_{1,1}\geq n_\tn{th}$, and if the verification is correct or not, i.e., $r_{1,1}/s_{1,1}\leq \delta_\tn{th}$.~However, in the case of weak laser sources, the direct observation of $s_{1,1}$ and $r_{1,1}$ is not possible as the verifiers do not know which of the successful BSM detections are due to single-photon emissions.~To overcome this issue, \cref{protocol:decoy} uses the decoy-state method as a means to construct random one-sided intervals for $s_{1,1}$ and $r_{1,1}$. In particular, the intervals $\rv{s}^\tn{lb}_{1,1}$ and $\rv{r}^\tn{ub}_{1,1}$, as specified in Eqs.~\eqref{Eq:decoy_1} and \eqref{Eq:decoy_2}, are constructed to capture $s_{1,1}$ and $r_{1,1}$ with very high probability in each run of the protocol.

The key point here is that although the decoy-state method can be made very reliable (i.e., by choosing a large $\nu$), there is still a non-vanishing probability that the intervals will fail to capture $s_{1,1}$ and $r_{1,1}$ in the right direction.~That is, there could be instances of the protocol in which the computed intervals are wrong and yet the verifiers agree with the adversaries.~In terms of the security analysis, this means that there is a strictly non-zero probability that the verifiers will agree with the adversaries, thereby implying an additional source of soundness errors.~Here, it is important to emphasize that this source of soundness error (which is due to the uncertainties in the decoy-state method) is fundamentally different from the soundness error captured by Eq.~\eqref{eq:soundness_qubit}, which is induced by the uncertainty in the error rate distribution.~Below, we show that the soundness error of \cref{protocol:decoy} is the same as \cref{protocol:qubit} except for an additive error term that is due to the statistical errors of the decoy-state method used.

\begin{namedthm}{Result}[Security with weak laser sources]\emph{Given $\{\mu_1, \mu_2, \mu_3\}$, $\{p_u \times p_v\}_{u,v}$,~$n_\tn{th}$,~$\delta_\tn{th}$,~and $\nu$, the protocol is $\varepsilon_\tn{decoy}$-sound with}
\be\label{eq:soundness_decoy}
\varepsilon_\tn{decoy} < \varepsilon_\tn{qubit}+ 2\eps_1 + \eps_2,\ee
where $\eps_1:=1-(1-e^{-2\nu})^7$ and $\eps_2:=1-(1-e^{-2\nu})^4$.
\end{namedthm}

\begin{proof}Here, we start from a general scenario and assume that the adversaries use $s_{1,1}>n_\tn{th}$ with probability $\kappa$ and $s_{1,1}\leq n_\tn{th}$ with probability $1-\kappa$.~Note that this choice of partitioning is not restrictive (since $\kappa$ is not fixed) and is merely used to faciliate the security analysis.~Let the event $s_{1,1}>n_\tn{th}$ be denoted by $\Theta$, then the soundness error can be written as
\[
\varepsilon_\tn{decoy}=1-\kappa\Pr\left[ \mathtt{N}|\Theta\right]-(1-\kappa)\Pr\left[ \mathtt{N}|\Theta^\tn{c}\right].
\] By conditioning on $\Omega_\tn{qc}$, we further get 
\begin{eqnarray} \nonumber
\varepsilon_\tn{decoy}&=&\kappa\Pr[\Omega_\tn{qc}|\Theta]\left(1-\Pr\left[ \mathtt{N}|\Theta,\Omega_\tn{qc}\right]\right) \\ \nonumber&&+(1-\kappa)\Pr[\Omega_\tn{qc}|\Theta^\tn{c}]\left(1-\Pr\left[ \mathtt{N}|\Theta^\tn{c},\Omega_\tn{qc}\right]\right).
\end{eqnarray}
The above can be simplified by setting $\Pr\left[ \mathtt{N}|\Theta^\tn{c},\Omega_\tn{qc}\right]=0$ and $\kappa$, $\Pr[\Omega_\tn{qc}|\Theta] \leq 1$ to get a bound that is independent of $\kappa$ (which is unknown), 
\be
\varepsilon_\tn{decoy}< 1-\Pr\left[ \mathtt{N}|\Theta,\Omega_\tn{qc}\right]+ \Pr[\Omega_\tn{qc}|\Theta^\tn{c}]. \label{eq:proof2_0}
\ee

Now, let us focus on the event $\Theta$, where there are two parts to it.~The first part consists in bounding the probability that $\rv{r}_{1,1}/s_{1,1}>\delta_\tn{th}$.~This is given by Eq.~\eqref{eq:soundness_qubit} with $n_\tn{th}$ replaced by $s_{1,1}$: $\Pr[\rv{r}_{1,1}/s_{1,1}>\delta_\tn{th}]> 1-\varepsilon'_\tn{qubit}$, where we used $\varepsilon'_\tn{qubit}$ to remind that $s_{1,1}$ has been used instead of $n_\tn{th}$.~Then from $\varepsilon'_\tn{qubit}<\varepsilon_\tn{qubit}$ , we have
\be \label{eq:proof2_a}
\Pr[\rv{r}_{1,1}/s_{1,1}>\delta_\tn{th}]> 1-\varepsilon_\tn{qubit}, \ee which is now expressed in terms of the protocol parameters. The second part consists in bounding the reliability of the decoy-state method.~Recall that the goal is to provide a lower bound on $s_{1,1}$ and an upper bound on $\rv{r}_{1,1}=r_{1,1}$ (i.e., for a given realization of $\rv{r}_{1,1}$).~These bounds are given by $\rv{s}_{1,1}^\tn{lb}$ and $\rv{r}_{1,1}^\tn{ub}$, which are one-sided interval estimates.~Suppose for the moment the reliability of these estimates are known, i.e.,~$\Pr[s_{1,1}\break > \rv{s}_{1,1}^\tn{lb}]>1-\eps_1$ and $\Pr[r_{1,1} < \rv{r}_{1,1}^{\tn{ub}}|\rv{r}_{1,1}=r_{1,1}]>1-\eps_2$.~Then, by taking the ratio distribution, we can construct an one-sided interval for the single-photon error rate,
\be\label{eq:proof2_b}
\Pr\!\left[{r_{1,1}}/{s_{1,1}} < {\rv{r}_{1,1}^\tn{ub}}/{\rv{s}_{1,1}^{\tn{lb}}}|\rv{r}_{1,1}=r_{1,1}\right] >(1-\eps_1)(1-\eps_2).
\ee
Operationally, this means that given $s_{1,1}$ and $r_{1,1}$, the decoy-state method \emph{will} output a single-photon error rate estimate, ${\rv{r}_{1,1}^\tn{ub}}/{\rv{s}_{1,1}^{\tn{lb}}}$, that is larger than the true single-photon error rate ${r_{1,1}}/{s_{1,1}}$ with probability greater than $(1-\eps_1)(1-\eps_2)$.~Notice that the probability statement is about the computed interval and not about the true single-photon error rate.

Now it remains to put everything together.~First, we have that the probability of rejection conditioned on $\Theta$ is given by
\[ 
\Pr\left[ \mathtt{N} |\Theta,\Omega_\tn{qc}\right]=\Pr[\delta_\tn{th}<{\rv{r}_{1,1}^\tn{ub}}/{\rv{s}_{1,1}^{\tn{lb}}}|\Theta,\Omega_\tn{qc}],
\] which is essentially  Eq.~\eqref{eq:proof2_b} conditioned on the event $\rv{r}_{1,1}>\lceil\delta_\tn{th} s_{1,1} \rceil$.~More precisely, we have $\Pr\left[ \mathtt{N} |\Theta,\Omega_\tn{qc}\right]=\Pr[\rv{r}_{1,1}>\lceil\delta_\tn{th} s_{1,1} \rceil|\Theta]\Pr[{r_{1,1}}/{s_{1,1}} < {\rv{r}_{1,1}^\tn{ub}}/{\rv{s}_{1,1}^{\tn{lb}}}|\rv{r}_{1,1}=r_{1,1}]$, which together with Eq.~\eqref{eq:proof2_a} implies
\[
\Pr\left[ \mathtt{N} |\Theta,\Omega_\tn{qc}\right]> (1-\eps_1)(1-\eps_2)(1-\varepsilon_\tn{qubit}).
\] Plugging this in Eq.~\eqref{eq:proof2_0}, we thus get
\begin{eqnarray*}
\varepsilon_\tn{decoy}&<& 1 - (1-\eps_1)(1-\eps_2) + \varepsilon_\tn{qubit}+ \Pr[\Omega_\tn{qc}|\Theta^\tn{c}]\\
&<&1 - (1-\eps_1)(1-\eps_2) + \varepsilon_\tn{qubit}+ \eps_1\\
&<& 2\eps_1+\eps_2 + \varepsilon_\tn{qubit},
\end{eqnarray*}
where in the second inequality we used $\Pr[\Omega_\tn{qc}|\Theta^\tn{c}] \leq \eps_1$.

Finally, in Appendix \ref{App:ss2} we show that the statistical errors $\eps_1$ and $\eps_2$ can be parameterized using a fixed security constant, $\nu$, giving 
\[
\eps_1=1-(1-e^{-2\nu})^7,\quad \eps_2=1-(1-e^{-2\nu})^4,\]
which concludes our proof sketch.
\end{proof}

\begin{figure}[t!]
\centerline{\includegraphics[width=7.6cm]{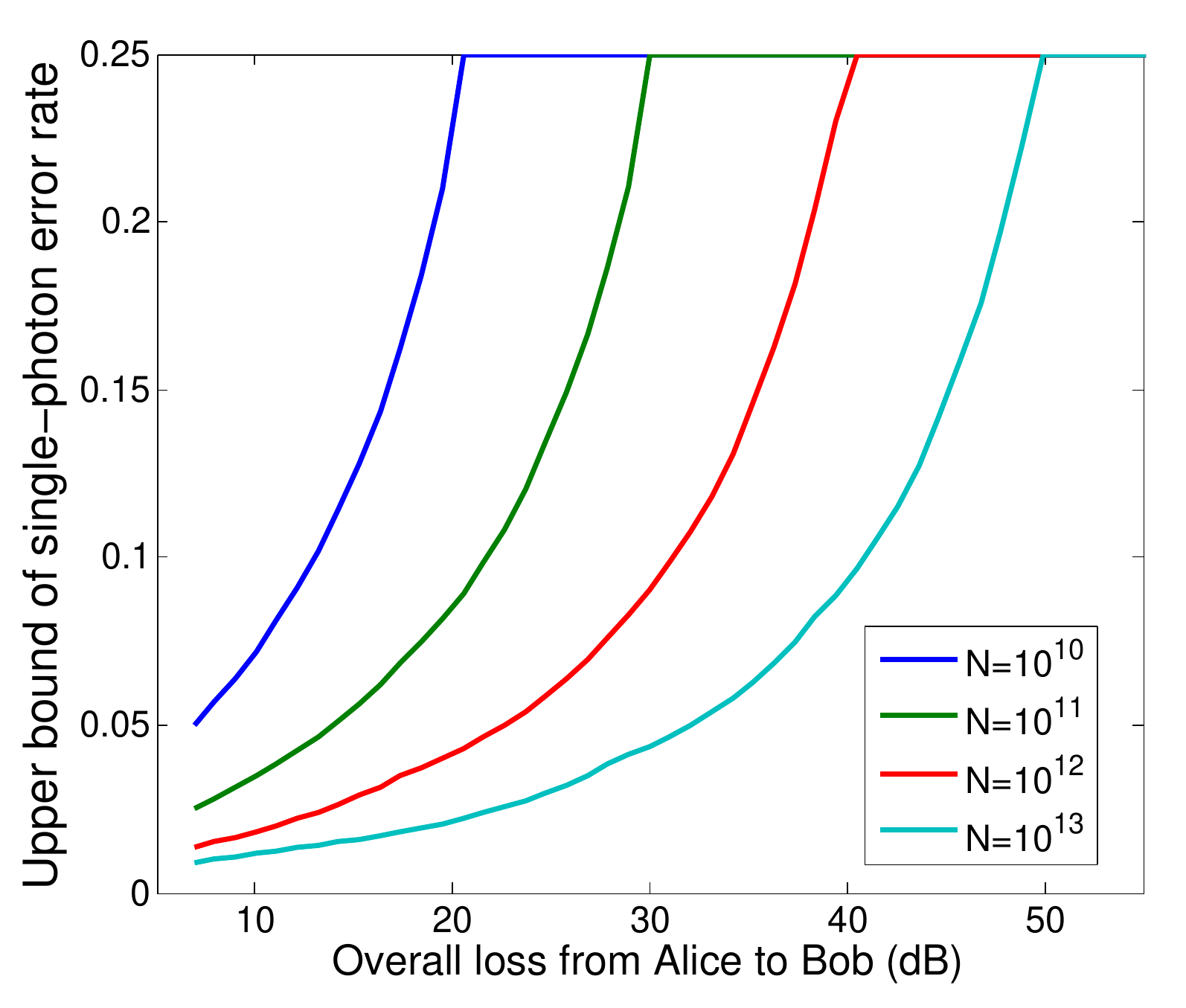}} \caption{\textbf{The upper bound of the estimated single-photon error rate versus overall loss between Alice and Bob.} The simulation assumes a baseline QBER of $0.1\%$. The the detectors are assumed to have an efficiency of 64\% and a dark count rate of  $2.5\times 10^{-6}$. The starting cut-off point is about 6.8 dB, which is the total loss in the BSM. The numerical results are obtained using $N=10^x$ with $x=10,11,12,13$ (from left to right).}\label{fig3:errorrate}
\end{figure}

A way to evaluate the feasibility of our protocol is to look for the loss point (in dB) at which the error rate, ${\rv{r}_{1,1}^\tn{ub}}/{\rv{s}_{1,1}^{\tn{lb}}}$, is greater than $1/4$.~To this end, we consider a symmetric photonic implementation where the prover is positioned at the center between the verifiers, i.e., see Fig.~\eqref{fig1:spacetimecone}.~The implementation is based on polarized photons, linear optical elements and threshold detectors.~Following standard channel error models for photonic quantum communication (e.g., see Ref.~\cite{Xu13}), we assume two sources of error, namely polarization misalignment errors and background noise.~In this case, the quantum bit error rate (QBER) is made up of two components: a baseline error rate (polarization misalignment errors) and a loss-dependent error rate (due to detector dark counts).~Evidently in our consideration, the limit on the amount of tolerable loss is largely determined by the detector dark count rate.~For the simulation, we borrow experimental parameters from a recent MDI-QKD experiment~\cite{Tang16}:~the baseline error rate is fixed to 0.1\% and the detectors (with $64\%$ efficiency) are assumed to have a dark count rate of $2.5\times10^{-6}$. Also, the security parameter of the decoy-state method to fixed to $\nu=10$, giving an overall error probability of $\sim 10^{-8}$. In Fig.~\eqref{fig3:errorrate}, we plot ${\rv{r}_{1,1}^\tn{ub}}/{\rv{s}_{1,1}^{\tn{lb}}}$ for $N=10^x$ with $x=10,11,12,13$ against the overall quantum channel loss (dB). From the simulation, we see that our protocol is able to tolerate up to about 47 dB loss with weak laser sources.

\section{Conclusion and outlook}\label{sec:dis}
In the above, we have presented a time-reversed entanglement swapping QPV protocol that is highly robust against detection losses.~Using a proof technique from Refs.~\cite{Cosentino2013,Lim2016}, we first showed that \cref{protocol:qubit} (assuming ideal BB84 qubits) is secure against arbitrary local operations and classical communication (LOCC) attacks for any quantum channel loss.~In particular, the soundness error of the protocol is shown to be independent of the overall detection loss and is exponentially small in the number of rounds with conclusive measurement outcomes.~This is in contrast to the widely studied BB84 QPV protocol, which is insecure when the quantum channel loss is $\geq 1/2$ assuming LOCC attacks~\cite{Qi15}.~In Section \ref{sec:4}, we extended \cref{protocol:qubit} to weak laser sources using a practical decoy-state method with three intensities (denoted by \cref{protocol:decoy}).~We found that the soundness error of \cref{protocol:decoy} only degrades by an additive error term that is dependent on the reliability of the underlying decoy-state method.~In addition, we performed numerical simulations using realistic experimental conditions and found that secure position verification is possible up to about 47 dB loss. 

Evidently, our proposed protocol is not the complete solution to practical QPV. In particular, what we have addressed here is only the overall detection loss assuming the verifiers are able to accurately prepare their quantum states.~To this end, it would be useful to investigate the impact of state preparation errors, especially considering the fact that such errors are known to severely degrade the security performance of quantum key distribution~\cite{Gottesman2004}.~One possible solution is to adopt the notion of \emph{loss-tolerant quantum cryptography}~\cite{Tamaki2014} and employ mismatched basis statistics to guarantee the loss-tolerant property of our protocol in the presence of state preparation errors.~Another interesting line of research would be to look at the semi-device-independent security of our protocol assuming that the dimensions of the verifier's quantum challenges (states) are fixed.~Several results have been obtained in this direction for measurement-device-independent QKD~\cite{Yin2013,Yin2014}, which suggests that similar conclusions could hold for our QPV protocol.

\acknowledgments{
This work was performed at Oak Ridge National Laboratory (ORNL), operated by UT-Battelle for the U.S. Department of Energy under Contract No. DE-AC05-00OR22725.~The authors acknowledge support from ORNL laboratory directed research and development program (LDRD), the U.S.~Department of Energy Cybersecurity for Energy Delivery Systems (CEDS) program program under contract M614000329, and the U.S. Office of Naval Research (ONR). }
\appendix

\section{Details of semidefinite program}
\subsection{SDP: preliminaries}
~In order for us to provide a more precise description of our semidefinite programs, we would need to introduce a few mathematical notations; some of which may be different from those used in the main text.~We let $V_1$ and $V_2$ complex Hilbert spaces be denoted by $\mathcal{A}$ and $\mathcal{B}$, respectively.~The set of linear operators, Hermitian operators and positive semidefinite operators acting on the composite Hilbert space are written as $\tn{L}(\mathcal{A}\otimes \mathcal{B})$, $\tn{Herm}(\mathcal{A}\otimes \mathcal{B})$ and $\tn{Pos}(\mathcal{A}\otimes \mathcal{B})$, respectively.~Furthermore, we write $Q \succeq 0$ to indicate that $Q$ is positive semidefinite.~The set of density operators corresponding to the verifiers' quantum systems is defined as $\tn{D}(\mathcal{A} \otimes \mathcal{B}):=\{\rho \in \tn{Pos}(\mathcal{A}\otimes \mathcal{B}) : \Tr[ \rho ]=1\}$.~Additionally, we would require the partial transpose operation, $T_\mathcal{B}=\mathbb{I}_{\tn{L}(\mathcal{A})} \otimes T$, which performs the transpose operation, $T$, on $V_2$'s Hilbert space.~Accordingly, the set of positive partial transpose (PPT) operators is defined as $\tn{PPT}(\mathcal{A} : \mathcal{B}):=\{Q:T_\mathcal{B}(Q) \succeq 0, Q \in \tn{Pos}(\mathcal{A}\otimes \mathcal{B}) \}$.~Also, we denote a diagonal matrix by $Q=\tn{diag}[\lambda_1, \lambda_2, \lambda_3,\lambda_4]$. 

\subsection{Optimal guessing probabilities}{\label{App:ss1}}~As mentioned in the main text, the bound for PPT measurements can be analytically solved using convex optimization techniques, namely, semidefinite programming~\cite{SDP1996}.~More specifically, the idea is to find feasible analytical solutions for the primal and dual programs, which provide lower and upper bounds on the optimal value (i.e., the weak duality principle).~If the solutions lead to values that coincide, then we say that the optimal solution for the semidefinite program is found.~That is, by the strong duality principle, the duality gap is zero.~In the following, we will show that the considered semidefinite programs have zero duality gaps.

\begin{namedthm}{Result}{\tn{\textbf{(Optimal guessing probability for PPT measurements).}}~The maximum probability of discriminating $\rho_0$ and $\rho_1$ using measurements $\{\Pi_0,\Pi_1,\Pi_\varnothing \} \in \tn{PPT}(\mathcal{A} \otimes \mathcal{B}) $ for any conclusive rate $\eta \in (0,1]$ is  }
\be \label{App:Eq:thm3}
P_\tn{guess}^{\tn{max}}( \cdot|\tn{PPT}) =\frac{3}{4}
\ee
\end{namedthm}
\begin{proof}The primal program for PPT measurements is given as \\

 \noindent 
\underline{{Primal program (PPT)}}
\begin{eqnarray*}\nonumber
\texttt{maximize}&:& \frac{1}{2}\Tr \left[\rho_0  \Pi_{0}  + \rho_1 \Pi_{1} \right] \\ \nonumber
\texttt{subject to}&:&  \Pi_{0}+ \Pi_{1}+ \Pi_{\varnothing} = \mathds{1}_{\mathcal{A} \otimes \mathcal{B}}\\  \nonumber
&& \Tr\left[ \rho_i\Pi_{\varnothing}\right] =1-\eta,\quad i=0,1\\
&& \Pi_k \in \tn{PPT}(\mathcal{A} :\mathcal{B}),\quad k=0,1,\varnothing,
\end{eqnarray*}and the corresponding dual program is \newline

\noindent
\underline{{Dual program (PPT)}}
\begin{eqnarray*}\nonumber
\texttt{minimize}&:& \Tr\left[Y \right] - (1-\eta)\gamma\\ \nonumber
\texttt{subject to}&:& 2\left(Y -T_{\mathcal{B}}(Q_i)\right)- \rho_i  \succeq 0,\quad i=0,1\\  \nonumber
&& 4\left(Y-T_{\mathcal{B}}(Q_2)\right) - \gamma\mathds{1}_{\tn{L}(\mathcal{A}\otimes \mathcal{B})} \succeq 0 \\ \nonumber
&& Y \in \tn{Herm}(\mathcal{A} \otimes \mathcal{B})\\
&& Q_i \in \tn{Pos}(\mathcal{A} \otimes \mathcal{B}),\quad i=0,1,2 \\
&& \gamma \in \mathbb{R}.
\end{eqnarray*} 

To prove Eq.~\eqref{App:Eq:thm3}, we need to construct feasible solutions for the primal and dual programs and show that their optimization values are identical.~For the primal program, a feasible solution is 
\begin{eqnarray*}
\tilde{\Pi}_{0}&=&\frac{1}{2}\begin{bmatrix}
    \eta &0 & 0 & 0 \\
    0 & \eta & \eta & 0 \\
    0 & \eta & \eta & 0 \\
    0 & 0 & 0 & \eta
  \end{bmatrix}, \quad  \tilde{\Pi}_{0}=\frac{1}{2}\begin{bmatrix}
    \eta &0 & 0 & 0 \\
    0 & \eta & -\eta & 0 \\
    0 & -\eta & \eta & 0 \\
    0 & 0 & 0 & \eta
  \end{bmatrix}\\ 
\tilde{\Pi}_{\varnothing}&=&\tn{diag}\left[1-\eta, 1-\eta,1-\eta,1-\eta \right].
\end{eqnarray*}Using this solution, we get $\eta P_\tn{guess}^{\tn{max}}( \eta|\tn{PPT}) \geq 3\eta/4$. For the dual program, a feasible solution is
\begin{eqnarray*}
\tilde{Y}&=&\frac{3}{16}\mathds{1}_{\tn{L}(\mathcal{A}\otimes \mathcal{B})},\quad \tilde{\gamma}=\frac{3}{4},  \\
Q_0&=&\frac{1}{16}\begin{bmatrix}
    1&0 & 0 & -1 \\
    0 & 0 & 0 & 0 \\
    0 & 0 & 0 & 0 \\
    -1 & 0 & 0 & 1
  \end{bmatrix}, \quad Q_1=\frac{1}{16}\begin{bmatrix}
    1&0 & 0 & 1 \\
    0 & 0 & 0 & 0 \\
    0 & 0 & 0 & 0 \\
    1 & 0 & 0 & 1
  \end{bmatrix},\\ Q_2&=&0_{\tn{L}(\mathcal{A}\otimes \mathcal{B})},
\end{eqnarray*} which gives $\eta P_\tn{guess}^{\tn{max}}( \eta|\tn{PPT}) \leq 3\eta/4$.~Putting everything together, the obtained optimal values give Eq.~\eqref{App:Eq:thm3}.
\end{proof}

\section{Details of decoy-state method}

Here, we provide the details for the bounds from the decoy-state analysis presented in the main text.~The analysis is mainly based on Ref.~\cite{Curty14}.

\subsection{Decoy-state method: preliminaries}

Our decoy-state method consists in both verifiers randomly setting the intensities of their respective laser pulses to one of the three intensity levels,~$\CMcal{I}=\{\mu_\rs,\mu_\rd,\mu_\rdd\}$ where $\mu_\rs > \mu_\rd+\mu_\rdd$ and $\mu_\rd>\mu_\rdd\geq 0$.~To analyze the finite-size effects of the decoy-state method, we consider an equivalent protocol, where $V_1$ ($V_2$) has the ability to send $k$-photon ($l$-photon) states, and they only decide on the choice of the average photon-number after the prover announces a successful measurement. In what follows, we will first introduce basic notations for the decoy-state analysis and then provide the relevant bounds for $s_{1,1}$ and $r_{1,1}$.

Let $s_{k,l}$~be the number of successful measurements announced by the prover given that $V_1$ has sent $k$-photon states and $V_2$ has sent $m$-photon states.~In this case, it is not hard to see that $\sum_{k,l=0}^\infty s_{k,l}=\sum_{u,v}n^{u,v}=n$ is the total number of detections, where $n_{u,v}$ is the number of detections assigned to intensity settings $u$ and $v$. Furthermore, we expect the size of $n_{u,v}$ to be
\be\label{A:eqn1}
\tilde{n}^{u,v}= \sum_{k,l=0}^\infty p_{u,v|k,l}s_{k,l},
\ee where $p_{u,v|k,l}$ is the conditional probability of choosing the intensity settings $u$ and $v$ given that $V_1$ sent a $k$-photon state and $V_2$ sent a $l$-photon state. More formally, the difference between the expected value ($\tilde{n}^{u,v}$) and the observed value ($n^{u,v}$) can be quantified by using the Hoeffding's inequality~\cite{hoeffding63}:
\be \label{A:eqn:1}
\left|\tilde{n}^{u,v}-n^{u,v} \right| < \Delta(n,\eps_1),
\ee
where $\Delta(n,\eps_1):=\sqrt{n/2\log(1/\eps_1)}$. The same statistical inequality can also be made for the expected number of errors and the observed number of errors for any pair of intensity settings.~Let $r_{k,l}$ be the number of errors associated with $s_{k,l}$, $m=\sum_{k,l=0}^\infty r_{k,l}$ be the total number of errors, and \be \label{A:eqn:3}
 \tilde{m}^{u,v}=\sum_{k,l=0}^\infty p_{u,v|k,l}r_{k,l}, 
\ee
be the expected number of errors assigned to intensity settings $u$ and $v$.~Then, the difference between $ \tilde{m}^{u,v}$ and $m^{u,v}$ is given by 
\be \label{A:eqn:2}
\left|\tilde{m}^{u,v}-m^{u,v} \right| < \Delta(m,\eps_2),
\ee 
which holds with probability at least $1-2\eps_2$. 

A central ingredient in Eqs. \eqref{A:eqn1} and \eqref{A:eqn:3} is the probability of choosing intensities $u,v$ given $k,l$ photons (i.e., $p_{u,v|k,l}$), which is not directly accessible in \cref{protocol:decoy}. To estimate this quantity, we note that with Bayes' rule, for all $u$ and $v$, we have
\be \label{A:eqn3}
p_{u,v|k,l}=\frac{p_{u,v}}{\tau_{k,l}}p_{k,l|u,v}=\frac{p_{u,v}}{\tau_{k,l}}\frac{e^{-(u+v)}u^k v^l}{k!l!},
\ee
where $p_{u,v}$ denotes the probability that $V_1$ chooses intensity $u$ and $V_2$ chooses intensity $v$, and
\be
\tau_{k,l}:=\sum_{u,v}p_{u,v}e^{-(u+v)}\frac{u^k v^l}{k!l!},
\ee
is the probability that $V_1$ prepares a $k$-photon state and $V_2$ prepares a $l$-photon state.\\

\subsection{Estimation of $s_{1,1}$ and $r_{1,1}$} \label{App:ss2}

Next, we discuss how to calculate $s_{1,1}$.~This is done by exploiting the structure of Eq.~\eqref{A:eqn1} and following the approach proposed by Refs.~\cite{Xu13,Curty14}. The estimation method is mainly based on Gaussian elimination.~For brevity, let $\xi^{u,v}:=\exp(u+v)p_u^{-1}p_v^{-1}$ for all $u,v\in \CMcal{I}$, then we have $s_{1,1} \geq s_{1,1}^\tn{lb}$ where
\be \label{Eq:lb_sp}
s_{1,1}^{\tn{lb}}=\left\lfloor\frac{(\mu_\rs^2-\mu_\rdd^2)(\mu_\rs-\mu_\rdd)\gamma_2' 
- (\mu_\rd^2-\mu_\rdd^2)(\mu_\rd-\mu_\rdd)\gamma_1'}{(\mu_\rs -\mu_\rdd)^2(\mu_\rd-\mu_\rdd)^2(\mu_\rs-\mu_\rd)}\right\rfloor,
\ee
and
\begin{multline}
\gamma_1':=\xi^{\mu_\rs,\mu_\rs}\tilde{n}^{\mu_\rs,\mu_\rs}+\xi^{\mu_\rdd,\mu_\rdd}\tilde{n}^{\mu_\rdd,\mu_\rdd}\\-\xi^{\mu_\rs,\mu_\rdd}\tilde{n}^{\mu_\rs,\mu_\rdd} -\xi^{\mu_\rdd,\mu_\rs}\tilde{n}^{\mu_\rdd,\mu_\rs}, \end{multline}
\begin{multline}
\gamma_2':=\xi^{\mu_\rd,\mu_\rd}\tilde{n}^{\mu_\rd,\mu_\rd}+\xi^{\mu_\rdd,\mu_\rdd}\tilde{n}^{\mu_\rdd,\mu_\rdd}\\-\xi^{\mu_\rd,\mu_\rdd}\tilde{n}^{\mu_\rd,\mu_\rdd}-\xi^{\mu_\rdd,\mu_\rd}\tilde{n}^{\mu_\rdd,\mu_\rd}.
\end{multline} 
An upper bound on the number of errors associated with the single-photon detection events is given in Refs~\cite{Xu13,Curty14}
\be \label{Eq:ub_errors}
r_{1,1}^{\tn{ub}}=\min \left\{ \left \lceil \frac{\gamma_3'}{(\mu_\rd-\mu_\rdd)^2} \right\rceil, \left \lceil \frac{s_{1,1}^\tn{lb}}{2} \right\rceil\right\}
\ee
where
\begin{multline}\gamma_3':= \xi^{\mu_\rd,\mu_\rd}\tilde{m}^{\mu_\rd,\mu_\rd}+\xi^{\mu_\rdd,\mu_\rdd}\tilde{m}^{\mu_\rdd,\mu_\rdd}\\-\xi^{\mu_\rd,\mu_\rdd}\tilde{m}^{\mu_\rd,\mu_\rdd}-\xi^{\mu_\rdd,\mu_\rd}\tilde{m}^{\mu_\rdd,\mu_\rd}.
\end{multline} At this point, Eqs.~\eqref{Eq:lb_sp} and~\eqref{Eq:ub_errors} are given in terms of $\tilde{n}^{u,v}$ and $\tilde{m}^{u,v}$, which are expected values. To rewrite the equations in terms of the observed values, we use Eqs.~\eqref{A:eqn:1} and~\eqref{A:eqn:2} to get 
\begin{eqnarray} \label{A:eqn:7}
 n^{u,v}- \sqrt{\nu n}& <&\tilde{n}^{u,v} < n^{u,v}+ \sqrt{\nu n}, \\ \label{A:eqn:8}
 m^{u,v}- \sqrt{\nu m} &<&\tilde{m}^{u,v} < m^{u,v}+ \sqrt{ \nu m}, 
\end{eqnarray} for all $u,v \in \CMcal{I}$.~Thus for a given security parameter $\nu>0$, the error probability for these inequalities is $\exp(-2\nu)$. In other words, each of the above inequalities holds with probability at least $1-\exp(-2\nu)$. Note that Eqs.~\eqref{Eq:lb_sp} and~\eqref{Eq:ub_errors} use 7 estimators and 4 estimators, respectively. 

Finally, by applying Eqs.~\eqref{A:eqn:7} and~\eqref{A:eqn:8} to Eqs.~\eqref{Eq:lb_sp} and~\eqref{Eq:ub_errors}, we arrive at the main equations for \cref{protocol:decoy},  Eqs.~\eqref{Eq:decoy_1} and~\eqref{Eq:decoy_2}.


\begin{thebibliography}{00}
\bibitem{footnote1}
In fact, it suffices to assume that the verifiers share an authenticated classical channel: they can perform sifting (as in QKD) to postselect measurement outcomes with matching bases.
\end{thebibliography}
\end{document}